\documentclass[10pt]{book}
\usepackage{array}
\usepackage{amssymb}
\usepackage{amsmath}
\begin{document}

%%%%%%%%%%%%%%%%%%%%%%%%%%%%%%%%%%%%%%%%%%%%%%%%%%%%%%%%%%%%%%%%%%%%%%%%%%%%%%%%%%%%%%%%%%%%%%%%%%%%%%%%%%%%%%%%%%%%%%%%%%%%
%%%%%%%%%%%%%%%%%%%%%%%%%%%%%%%%%%%%%%%%%%%%%%%%%%%%%%%%%%%%%%%%%%%%%%%%%%%%%%%%%%%%%%%%%%%%%%%%%%%%%%%%%%%%%%%%%%%%%%%%%%%%

\renewcommand{\baselinestretch}{1.2}

\markright{
\hbox{\footnotesize\rm Statistica Sinica (2013): Preprint}\hfill
}

\markboth{\hfill{\footnotesize\rm K. Chen, Y. Lin, Y. Yao AND C. Zhou} \hfill}
{\hfill {\footnotesize\rm Regression analysis with response-biased sampling} \hfill}

\renewcommand{\thefootnote}{}
$\ $\par

%%%%%%%%%%%%%%%%%%%%%%%%%%%%%%%%%%%%%%%%%%%%%%%%%%%%%%%%%%%%%%%%%%%%%%%%%%%%%%%%%%%%%%%%%%%%%%%%%%%%%%%%%%%%%%%%%%%%%%%%%%%%

\fontsize{10.95}{14pt plus.8pt minus .6pt}\selectfont
\vspace{0.8pc}
\centerline{\large\bf Regression Analysis with Response-biased Sampling}
\vspace{2pt}
\centerline{\large\bf }
\vspace{.4cm}
%\centerline{Kani Chen}
%\vspace{.4cm}
%\centerline{\it Affiliation(s)}
\centerline{Kani Chen}
\centerline{\it Department of Mathematics, Hong Kong University of Science
and Technology, Hong Kong}
%\centerline{\medit Email: makchen@ust.hk}
\centerline{Yuanyuan Lin}
\centerline{\it Department of Applied Mathematics, Hong Kong Polytechnic University, Hong Kong}
%\centerline{\medit Xiamen University, Xiamen, China}
%\centerline{\medit E-mail: linyy@xmu.edu.cn}
\centerline{Yuan Yao}
\centerline{\it Department of Mathematics, Hong Kong Baptist University, Hong Kong}
%\centerline{\medit Email: yaoyuan@ust.hk}
\centerline{Chaoxu Zhou}
\centerline{\it Department of Mathematics, Hong Kong University of Science
and Technology, Hong Kong}
%\centerline{\medit Email: cxzhou@ust.hk}
\vspace{.55cm}
\fontsize{9}{11.5pt plus.8pt minus .6pt}\selectfont

%%%%%%%%%%%%%%%%%%%%%%%%%%%%%%%%%%%%%%%%%%%%%%%%%%%%%%%%%%%%%%%%%%%%%%%%%%%%%%%%%%%%%%%%%%%%%%%%%%%%%%%%%%%%%%%%%%%%%%%%%%%%

\begin{quotation}
\noindent {\it Abstract:}
Response-biased sampling, in which samples
 are drawn from a population according to the
 values of the response variable, is common in biomedical, epidemiological, economic and social studies.
In particular, the complete observations in data with censoring, truncation or
missing covariates can be regarded as response-biased sampling under certain conditions.
This paper proposes to use transformation models, known as the generalized accelerated failure time model in econometrics,
for regression analysis with
response-biased sampling.
%The only finding
%in the litirature that equals is case-control logistic regression.
With unknown error distribution, the transformation models are broad enough to
cover linear regression models, the Cox's model and the proportional odds model as special cases.
To the best of our knowledge, except for the case-control logistic regression, there is no report in the literature that a prospective estimation approach can work for biased sampling without any modification.
We prove that the maximum rank correlation estimation is valid for response-biased sampling and
establish its consistency and asymptotic normality.
Unlike the inverse probability methods,
the proposed  method of estimation does not involve the sampling probabilities, which are often
difficult to obtain in practice.
Without the need of estimating the unknown transformation function or the error distribution,
the proposed method is
numerically easy to implement
with the Nelder-Mead simplex algorithm, which does not require convexity or continuity.
We propose an inference procedure using
 random weighting  to avoid the complication of density estimation when using the plug-in rule
 for variance estimation.
Numerical studies with supportive evidence are presented.
Applications are illustrated with the Forbes Global 2000 data
 and the Stanford heart transplant data.\par

\vspace{9pt}
\par
\end{quotation}\par
%\noindent \hskip .2in {\it\small Address for correspondence:} {\small Yuan Yao,
%Department of Mathematics,
%Hong Kong Baptist University, Hong Kong.}
%{\small E-mail: yaoyuan@hkbu.edu.hk}
%\par
%%%%%%%%%%%%%%%%%%%%%%%%%%%%%%%%%%%%%%%%%%%%%%%%%%%%%%%%%%%%%%%%%%%%%%%%%%%%%%%%%%%%%%%%%%%%%%%%%%%%%%%%%%%%%%%%%%%%%%%%%%%%

\fontsize{10.95}{14pt plus.8pt minus .6pt}\selectfont

\setcounter{chapter}{1}
\setcounter{equation}{0} %-1
\noindent {\bf 1. Introduction}

Response-biased sampling is commonly used in biomedical, epidemiological, financial and social studies.
In response-biased sampling,   observations are taken  according to the values of the responses.
Specifically, let $(Y^*, X^*)$ and $(Y, X)$ represent the pair of response and covariates
in the population and in the sample, respectively. In a response biased
sampling, the conditional distribution of $X$ given $Y$ is the same as that of $X^*$ given $Y^*$.
Throughout the paper, we denote the observations as $(Y_i, X_i), i=1,...,n$, which are independent
and identically distributed.
%As a result, the conditional distribution of the covariates given the response
%is the same
%as that in the population.
Data collected using response-biased sampling schemes are likely to contain more information
 relevant to one's interest than using prospective sampling.
Such retrospective sampling is useful in clinical studies
 for its effectiveness and its saving duration and costs.
For example, in a study of possible dependence of levels of hypertension (response) on
those of sodium intake (covariate),  sampling
from patients in a hospital, which can be regarded as
response-biased sampling,  would be more effective than from general public as the latter
 has much smaller proportion of people with hypertension.
 Moreover, a typical example of sampling with selection bias in economic and social studies is that the wage is only observed for the employed people.
 %The response-biased sampling that we consider can be viewed as a special case of the celebrated Heckman model; see Heckman (1977, 1979). The Heckman model assumes an outcome linear regression model and a probit selection model. We consider more general transformation models and assume the "selectivity/observability" solely depends on the value of the response variable. In the case analysis of wage, we assume the chance that a potential job is taken only depends on the wage offered. The proposed estimating method does not depend on the specification of the sampling probabilities, unlike the well known Heckman correction.
The statistical analysis of biased sampling has received considerable attention
in the past decades. Case-control or choice-based sampling, which
is a special case of response-biased sampling, has been extensively studied in the literature;
see Anderson (1972), Manski and Lerman (1977), Prentice and Pyke (1979),
Breslow and Day (1980), Cosslet (1981), Scott and Wild (1986, 1997), Manski (1993), etc.
There are other studies on biased sampling data, involving semiparametric and parametric
models; see Hausman and Wise (1981), Jewell (1985), Bickel and Ritov (1991), Wang (1996),
 Lawless et al. (1999),
Chen (2001), Tsai (2009), Luo and Tsai (2009), Luo et al. (2009), among others.
In statistical analysis of biased sampling, one of the celebrated findings is that
the prospective estimating equation is still valid for case-control logistic
regression; see Anderson(1977) and Prentice and Pyke (1979). However,
in general, estimating equations based on prospective sampling
will be invalid for biased sampling and modifications using, for example,
inverse probability method is necessary.
This paper shows, for general transformation model,
 a rank estimation method based on prospective sampling
still applies, without any modification, to response-biased sampling.

Regression analysis with response-biased sampling is generally associated with the fitted model.
In particular, the estimation of the parameter of interest with biased sampling usually
relies  on the model assumptions, such as the inverse probability   method  and
the pseudo-likelihood method; see Binder (1992), Lin (2000), Wang (1996) and Tsai (2009).
 Recently, nonparametric tests and estimation for right censored data with biased sampling can
be found in Ning, Qin and Shen (2010) and Huang and Qin (2011). Moreover, a novel approach to
analyze length-biased data with semiparametric transformation and accelerated failure time models has
been developed by Shen, Ning and Qin (2009).
In this paper, we consider a class of  transformation models with response-biased sampling,
under which an unknown monotonic transformation of the response is linearly related to the covariates
with an unspecified error distribution. The transformation models are also called the
generalized accelerated failure time (GAFT) model in econometrics.
This class of regression models includes many popular models,
such as the proportional hazards model, the proportional odds model
as well as  accelerated failure time models or linear models.
Furthermore,
the response-biased sampling that we consider can be viewed as a special case of the celebrated Heckman model; see Heckman (1977, 1979). The Heckman model assumes an outcome linear regression model and a probit selection model. We consider more general transformation models and assume the "selectivity/observability" solely depends on the value of the response variable. In the case analysis of wage, we assume the chance that a potential job is taken only depends on the wage offered.
The proposed estimating method does not depend on the specification of the sampling probabilities, unlike the well known Heckman correction.
We note that there is a rich literature on
 linear transformation models with a known error
distribution; see, for example,
Dabrowska and Doksum (1988), Cheng et al. (1995, 1997), Chen et al. (2002) and Zeng and Lin (2007).
However, their reported methods cannot be directly applied to transformation models with an unknown
error distribution.
Similarly, the case-control logistic regression method in Anderson (1977) and Prentice and Pyke (1979)
which works only for a special model, cannot be generalized directly and modification using, for
example, the inverse probability method is inevitable. However,
for the inverse probability method, its validity requires correct prospective mean zero estimating equations, correct specifications of sampling probabilities, and the sampling probability must be positive for every value in the range of the response.

% Some further analysis involving semiparametric regression models may be seen in
% Hausman and Wise (1981), Jewell (1985) and Bickel and Ritov (1991), among others. We note
% that the above stratified response-biased sampling and the response-biased sampling considered in
% this paper are different from another type of sampling characterized by missing covariates, such
% as those considered in, for example, Robins {\it et. al} (1994) and Lawless  {\it et. al} (1999)
% or the case-cohort and nested case-control sampling studied in, for example, Thomas (1977),
% Prentice (1986), Samuelsen (1997) and Chen and Lo (1999).
% Chen (2001) derived the semiparametric maximum likelihood estimate for the regression parameter
% in parametric models with response-biased sampling.

%??Check literature for recent development for response-biased sampling??

In view of the importance of the response-biased sampling designs as well as  transformation models,
an easy-to-implement estimation methodology, with an advantage over the
existing methods in terms of generality,  is worth pursuing.  Note that the conventional
methods, such as the least squares (LS) or
the least absolute deviations (LAD) cannot be directly applied
to response-biased sampling,
because the zero mean and the zero median assumptions do not hold anymore.
The maximum rank correlation (MRC) estimate, originated from Han (1987)
 for prospective studies, is based on the rank correlation (Kendall's $\tau$) between two variables.
For illustration, consider a simple linear regression model
$$Y_i=\beta'X_i+\epsilon_i, \ \ 1\le i\le n, $$
where
$(Y_i,X_i, \epsilon_i)$ are independent and identically distributed ($i.i.d.$) copies of $(Y,X,\epsilon)$.
The idea of the MRC estimation is to maximize the rank correlation between $Y_i$ and $\beta'X_i$ with respect to $\beta$. Heuristically, given that $\beta'X_i>\beta'X_j$, it is more likely that $Y_i>Y_j$
than otherwise. In other words, the rank of $Y_i$ and the rank of $\beta'X_i$ are positively correlated.
%In fact,
%the MRC estimate is motivated by such a simple principle.
A number of studies on MRC have been conducted.
Sherman (1993) proved its $\sqrt{n}$-consistency and asymptotic normality and Khan and Tamer (2004) extended this method to semiparametric models with censoring by proposing the partial rank (PR) estimator. A smoothed partial rank (SPR) estimator is then considered in Song et al. (2007) for transformation models with censoring.

Inspired  by the fact that response-biased sampling would not change the positive correlation between the ranks of the responses and explanatory variables, this article offers an estimator based on
MRC for transformation models with response-biased sampling.
%There are several advantages to this method:\\
%(a) Most importantly,
The proposed estimation does not rely on any further model assumption.
It works equally well regardless of what the monotonic transformation is, as the MRC estimate only
depends on the ranks of responses.
The estimation of the transformation function,
which is likely to be quite complex and computationally burdensome, is not required.
The proposed method is easy to implement and computationally straightforward with the help of
Nelder-Mead simplex direct search. It is quite well known that the Nelder-Mead simplex algorithm does not require convexity or continuity; see Nelder and Mead (1965).
Note that the MRC objective function is a U-statistic. In order to avoid
estimating the covariance matrix,  we propose to use a random weighting resampling scheme
for inference.
%(c) Neither parametric assumptions nor estimation of the error distribution is needed in our method.\\
In addition, since prospective sampling can be regarded as a special case
of response-biased sampling, the proposed estimation
is valid for prospective sampling.

We describe the model in section 2. The proposed estimation and its inference with theoretical
justification are presented in section 3. A simulation study with supportive evidence is given in section 4. In section 5, our method is applied to
the Forbes Global 2000 data set and the Stanford heart transplant data set.  The paper concludes with a remark in section 6. All proofs are deferred to the
Appendix.
\par

\setcounter{chapter}{2}
\setcounter{equation}{0} %-1
\noindent {\bf 2. Model description}

Let $(Y^*,W^*)$ be a pair of a response and a $(d+1)$-dimensional vector of covariates in the population. Assume that the response depends on the covariates according to the transformation model
\begin{eqnarray}
\label{1-0} H(Y^*)=\theta'_0W^* +\epsilon^*,
\end{eqnarray}
where $H(\cdot)$ is an unknown monotonically increasing   function,   $\epsilon^*$ is the error,
 independent of $W^*$, with unspecified distribution, and
$\theta_0$ is a $(d+1)$-dimensional vector of regression coefficients.
When $H(\cdot)$ is the identity function, model (\ref{1-0}) becomes a linear
regression model.
When $\epsilon^*$ follows the extreme-value distribution and the standard logistic distribution,
the resulting model is the proportional hazards model and the proportional odds model, respectively.
In prospective studies, model (\ref{1-0}) has been extensively studied in the literature.
Note that $\theta_0$ in model (\ref{1-0}) is not identifiable, meaning that $\theta_0$ is not uniquely
defined. To avoid unidentifiability, one may restrict   $\Vert \theta_0 \Vert = 1$.
 Without loss of generality, we choose to
 fix the first component of $\theta_0$ to be 1.
 Then,   $\theta_0=(1,\beta'_0)'$, where $\beta_0$ denotes the rest components.
%This condition
%is imposed as an identifiable condition, which is similar to assume the Euclidean norm of $\theta_0$
%is equal to 1.
Accordingly, $W^*$ can be decomposed into $W^*=(Z^*,X^*)$, where $Z^*$ is the covariate corresponding to the fixed regression coefficient and $X^*$ is the other $d$-dimensional covariate.
Hence, model (\ref{1-0}) can be rewritten as
 \begin{eqnarray}
\label{1-1} H(Y^*)=Z^*+\beta'_0X^* +\epsilon^*.
\end{eqnarray}

%Note that the transformation model is not identifiable as it is location and scale invariant. To avoid this problem, we fix the first component of $\theta$ to be 1 without loss of generality, that is $\theta_0=(1,\beta'_0)'$. Then $W^*$ can be decomposed as $W^*=(Z^*,X^*)$, where $Z^*$ is the %covariate corresponding to the fixed regression coefficient and $X^*$ is the $p$ dimensional covariates corresponding to the others. Hence, the original model can be rewritten as

Let $(Y, Z, X)$ be the response and covariates following the distribution of  response-biased sampling.
The nature of response-biased sampling implies that, for any $y$,
the conditional distribution of $(X,Z)$ given $Y=y$ is the same as that of $(X^*,Z^*)$ given $Y^*=y$.
An alternative but equivalent definition of response biased sampling is by using a
sampling index $\Delta$.      The pair of response
and
covariates,  $(Y^*, X^*)$,  are observed if and only if $\Delta=1$. Then the  response biased sampling
is defined by the conditional independence of $\Delta$ and $X^*$ given $Y^*$.
And we can denote the observations as $(Y^*_i, X_i^*, \Delta_i)$ where $\Delta_i =1$ for $i=1,..., n$.
With biased-sampling,
Wang (1996) provided a novel pseudo-likelihood method for Cox's proportional hazards model.
Recently, a pseudo-partial likelihood approach can be found in Tsai (2009). The existing methods for Cox's
model with biased-sampling are conceptually appealing and have clear interpretation.
To the best of our knowledge, no specific construction of regression analysis based on
transformation models with response-biased sampling is available in the literature.
In the next section, we propose a general estimation and inference procedure based on MRC
for model (\ref{1-1}) with response-biased sampling.
\par

\setcounter{chapter}{3}
\setcounter{equation}{0} %-1
\noindent {\bf 3. Estimation and inference}

With response-biased sampling, the observations are $(Y_i, Z_i, X_i)$, $1\le i\le n$,
which are $i.i.d.$ copies of $(Y, Z, X)$. Throughout the paper, $I(\cdot)$ is the indicator function.
Similar to  Han (1987),
the rank correlation for response-biased sampling is defined as
\begin{eqnarray}
\label{1-3} U_n(\beta)=&\sum\limits
_{i\neq j}I(Z_i+\beta'X_i> Z_j+ \beta'X_j)I(Y_i>Y_j).
\end{eqnarray}
The MRC estimate is to maximize the rank correlation
$U_n(\beta)$. Denote $\hat \beta_n$ as the maximizer of $U_n(\beta)$.
Han (1987) and Sherman (1993) established the consistency and asymptotic normality
of $\hat\beta_n$ with data from prospective sampling.
However, with response sampling,
it is not clear whether the large sample properties still hold.

The following theorem presents the consistency and asymptotic normality for $\hat \beta_n$ with
response-biased sampling.

{\bf Theorem 1.} {\it Under regularity conditions C1-C4 given in the Appendix, as $n \rightarrow \infty$,
$$\sqrt{n}(\hat\beta_n-\beta_0)\rightarrow N\left(0,A^{-1}B(A^{-1})' \right)$$ in distribution,
 where the explicit forms of $A$ and $B$ are given in the Appendix.}

It is shown in the Appendix that the limiting covariance matrix of $\hat{\beta}_n$ involves the derivative of
conditional expectation of the objective function,
which could be quite difficult to estimate. To circumvent the difficulty, we propose
a distributional approximation based on random weighting method by externally
generating $i.i.d.$ random variables. Let $e_1, \cdots, e_n$ be a sequence of
$i.i.d.$ nonnegative random variables with mean 1 and variance 1.
Define
\begin{eqnarray}
\label{1-3} \tilde U_n(\beta)=\sum_{i\neq j} e_ie_j I(Z_i+\beta'X_i>Z_j+ \beta'X_j)I(Y_i>Y_j)
\end{eqnarray}
and $\tilde \beta_n=\mbox{arg max}_{\beta\in \mathcal{B}} \tilde{U_n}(\beta)$. The distribution
of $\sqrt{n}(\hat\beta_n-\beta_0)$ can be approximated by the resampling distribution of
$\sqrt{n}(\tilde \beta_n-\hat\beta_n)$ when fixing the data $(Y_i, Z_i, X_i)$, $1\leq i \leq n$.
%Let
%$\mathcal{L}$ denote the conditional distribution given $\{(Y_i, Z_i, X_i), 1\leq i \leq n\}$.

{\bf Proposition.} {\it Given $\{(Y_i, Z_i, X_i), 1\leq i \leq n\}$, under regularity conditions C1-C4 in the Appendix, as  $n \rightarrow \infty$,
$$\sqrt{n}(\tilde{\beta}_n-\hat{\beta}_n)\rightarrow N\left(0,A^{-1}B(A^{-1})' \right)$$
in distribution, which is the asymptotic distribution of $\sqrt{n}(\hat\beta_n-\beta_0)$. }

The resampling method based on random weighting for the U-statistic objective function is well established in Jin (2001). We omit the proofs of the proposition here.

 \textit{Remark 1.}
For the computation, the numerical minimization
is straightforward with the Nelder-Mead simplex algorithm which does not require convexity or continuity. In the simulation, we use Nelder-Mead algorithm directly to search over a wide range of starting values in case there may exist local maximizers. Matlab code is available upon request. In addition, another slight problem is that, with large sample size or large dimension of covariates, the computation tends to be slower in simulation
due to many replications. However, an algorithm proposed by Abrevaya (1999) which improves the complexity of computation for MRC from $O(n^2)$ to $O(n\log n)$ is available for large sample size.
%Note that for large sample size, Abrevaya (1999) proposed an algorithm to improve the complexity of computation for MRC from $O(n^2)$ to $O(n\log n)$.
%Hence, the computational speed can be improved significantly.
%And for large dimension of covariates,
%the computational issue has been addressed in
And a smoothed approximation of the indicator function considered by Song et al. (2007)
can be applied for large dimension of covariates. Overall, the proposed method has little difficulty in numerical implementation.

 \textit{Remark 2.}
Note that our objective function $U_n(\beta)$ only depends on the responses through their orders which are not changed by
the unknown monotonically increasing transformation $H(\cdot)$.
Thus our estimate of $\beta_0$ is invariant of the transformation and estimating the unknown transformation $H(\cdot)$ can be avoided.

\textit{Remark 3.}
Response-biased sampling is related to truncated   and censored data.
With the presence of left-truncation, let $X^*$ be  covariates, $Y^*$ be response and $C^*$ be the left-truncation variable. Then,   $(X^*,Y^*,C^*)$ is observed if and only if $Y^*\geq C^*$,
and the observation, denoted as $(X, Y, C)$ accordingly, follows the conditional distribution
of $(X^*,Y^*,C^*)$ given $Y^*\geq C^*$.
%Denote their density functions as $f_{y^*}$, $f_{c^*}$ respectively.
The observed pair of covariates and response,
 $(X, Y)$, can be treated as a special case of response-biased sampling, if
 $C^*$ is independent of $X^*$ and $Y^*$. Specifically,   the conditional density
 of $X$ given $Y$ can be formally written as
$$
f_{X|Y}(x|y)
= \frac{f_{(X,Y)}(x,y)}{f_Y(y)}=\frac{\int f_{(X,Y,C)}(x,y,c)dc}{\int f_{(Y,C)}(y,c) dc}
 = \frac{\int f_{(Y, C)}(y, c)   f_{X|(Y,C)}(x|(y,c)) dc}{\int f_{(Y, C)}(y, c)   dc}.
$$
 The independence of $C^*$ and $(X^*,Y^*)$ gives $$f_{X|(Y,C)}(x|(y,c))=\frac{f_{(X^*,Y^*)}(x,y)f_{C^*}(c)/P(Y^*\geq C^* )}{f_{Y^*}(y)f_{C^*}(c)/P(Y^*\geq C^*)}=f_{X^*|Y^*}(x|y),$$
 which is irrelevant with $c$.
Thus, the conditional distribution of $X|Y$ is the same as that of $X^*|Y^*$.
Similarly, for right-censored  data with the censoring variable $
\tilde C$ independent of $X^*$ and $Y^*$, denote the observation   as $(X,Y,\delta)$, where
  $X=X^*$, $Y=\min(Y^*, \tilde C)$ and  $\delta= I(Y^*\leq
\tilde C)$. Then, the conditional density is
$$f_{X|(Y,\delta=1)}(x|(y,\delta=1))=\frac{f_{(X^*,Y^*)}(x,y)
P(\tilde C\geq y)}{f_{Y^*}(y)P(\tilde C \geq y)}=f_{X^*|Y^*}(x|y).$$
Therefore the uncensored observations can be regarded as drawn from a response-biased sampling. Note that the
partial rank method, which works for censoring data set, cannot be applied to truncation data.
The our method works better in this view as it can handle a broad class of data types including left-truncation and right-censoring.
%$Remark.$ Noting that our objective function $U_n(\beta)$ only depends on the responses through their orders, and the unknown monotone increasing transformation $H(\cdot)$ does not change the orders of the responses,  thus our estimate of $\beta_0$ is invariant of the transformation so that estimating the unknown transformation $H(\cdot)$ is also avoided. More importantly, we can get a more efficient estimation through retrospective sampling, empirical evidence is provided in next section.

\textit{Remark 4.} For data with missing covariates, the complete observations can be regarded as
drawn from a response biased sampling, if the missing mechanism is missing-at-random.
This is because, by the
  definition of missing-at-random,
 the conditional distribution of the
 covariates given the  response  for the complete cases
 is the same as that for the observations with missing covariates and, as a result,
 also same as that in the population.

\textit{Remark 5.}
Conditions C3 in the Appendix is imposed to
facilitate the proof of consistency.
We assume that the error distribution has a twice differentiable density function
with log-concavity. Although it looks somewhat restrictive,
it includes a number of widely used distributions, for example,  $N(0,\sigma^2)$ and Pareto family.
Thus linear models with normal errors, Cox's model and the proportional odds model are included.
With increasing technicalities, this condition might be loosened or dropped, as evidenced in
our simulation results in section 4.
\par

\setcounter{chapter}{4}
\setcounter{equation}{0} %-1
\noindent {\bf 4. Simulation studies}

Extensive simulation studies are conducted to examine the finite sample performance of the proposed
method, which are presented in four parts.
 In the first part, we consider  the linear model
\begin{eqnarray}
\label{m2}
 Y= Z+X_1\beta_1+X_2\beta_2+\epsilon,
\end{eqnarray}
where $(\beta_1,\beta_2)=(1, -1)$,  $Z\sim N(0,1)$, and $X_1$ and $X_2$  follow a bivariate normal distribution with mean $(1,-0.5)$ and
 variance
$$\begin{pmatrix} 1 & 0.2 \cr 0.2 & 1
\end{pmatrix}. $$
%The first set provides an evidence that parameter estimation in retrospective sampling is more efficient than that of in prospective sampling. The other set provides evidences that our method works well under different retrospective sampling schemes and different error distributions.
%The discontinuous nature of the objective function imposes difficulties in searching its maximum since the traditional gradient search methods such as Newton-Raphson algorithm cannot be applied. To tackle this problem, we use simplex optimization algorithms based method like the Nelder-Mead method instead.
%Another concern is the existing algorithms always converge to a local minimum rather than the global one. We can solve this problem by giving initial values in a large range and choose the optimal one by comparing their corresponding objective function values.
  Response-biased sampling are conducted with the five different schemes.
  In schemes 1, 2 and 5, the samples are restricted to
  $Y<-2$ or $Y>4$,  $Y>2.5$ and $3.8<Y<4.2$, respectively.
  In scheme 3, the sampling probability is fixed as   $\Phi(y-2)$ for response value $y$.
  Scheme 4 is simply the prospective sampling.
Four distributions for the error   are used:
 double exponential distribution with parameter 1,
the standard normal distribution, the standard
 extreme value distribution and the standard
 logistic distribution.
The sample size is  200 and simulation results are based on 100 replications.
 The external  random weights are generated from standard
exponential distribution with 500 replications.
For comparison, we also conduct simulation studies using inverse probability method with the same settings of the above five sampling schemes, in which the inverse sampling probabilities are the weights of the least square estimating equations; see Horvitz and Thompson (1952).
%For inference we set the resampling times to be 500 and the externally generated weighted random variables following exponential distribution with parameter $1$.
In Table 1, we present the bias of the estimates of the regression parameters $\beta_{1}$ and $\beta_{2}$ (BIAS), the empirical
standard error (SE), the average of the estimated standard errors (SEE) and
the $95\%$ coverage probabilities (CP) with the proposed method.
We also present the estimation results with the inverse probability method in Table 1.
\begin{center}
INSERT TABLE 1 HERE
\end{center}
%the simulation result in estimating $\beta_1$ and $\beta_2$ including the sample bias (Bias), the standard errors (SE) of the estimates,
%the estimated standard errors (SEE) and the $95\%$ coverage probabilities (CP) for $\beta_{1}$ and $\beta_{2}$.

It can be seen from Table 1 that the proposed method works well with all different sampling schemes and error distributions.
The estimated standard errors based on random weighting are close to the empirical standard errors in general.
The proposed method offers more accurate and stable estimates compared
with the inverse probability method for most of the examples. Except for sampling scheme 4
(prospective sampling) with normal and double exponential error
distributions, the inverse probability method gives inaccurate estimates in general. This is mainly
because the validity of the inverse probability method requires  correctly specifying
prospective mean zero estimating functions and positive sampling
probabilities, that are easily violated. Overall, the first part of the simulation contains strong evidence of the
superiority of the proposed method over the inverse probability method, in terms of both
generality and flexibility.

The second part of the simulation is intended to show
condition C3 may just be technical.
    Consider the model
$$Y  = Z + \beta'X  +\epsilon, $$
where $ \beta= 1$, $Z  \sim N(0,1)$, $X  \sim N(0,1)$, $\epsilon $ follows the mixture of
the standard normal distribution
and a Bernoulli distribution with probability of success $0.5$ and the mixture
probabilities are (0.5,0.5).  The error distribution is not
log-concave and thus  does not satisfy condition C3.
Samples with    values of the response  less than $-1.5$ or greater than $2.5$
are drawn.
 %The sample size is ???? and the simulation is replicated  100 times.
The bias of the estimate is 0.0302.
The empirical and estimated standard deviations are
  0.1591 and 0.1479, respectively.
The proposed method may still work without assuming
the log-concavity of the error distribution.

The third part uses a rather extreme example to demonstrate that
a biased sampling could be much more efficient than  prospective sampling.
Consider the model
\[Y=Z+\beta'X+\epsilon,\]
where $\beta=1$, $\epsilon \sim N(0,10^{-4})$,  $Z \sim N(0,1)$, and  $X$ follows
a distribution with  density function
\[ f_X(x) = \left\{ \begin{array}{ll}
         5*10^{-5},  & \mbox{if $-105 \leq x \leq -5 $};\\
         9.99, & \mbox{if $-0.05 \leq x \leq 0.05 $};\\
         5*10^{-5},  & \mbox{ if $5 \leq x \leq 105 $}.\end{array} \right. \]
For a response-biased sampling which only takes observations with responses
larger than $4.5$ or smaller than $-4.5$, the mean and standard deviation of
the estimates of $\beta$ are $1.0004$ and $0.0038$, respectively.
On the other hand, for the prospective sampling, the mean and
standard deviation of the estimates of $\beta$  are  $1.0108$ and $0.0576$, respectively.
 The relative efficiency for the response-biased sampling versus
 the prospective sampling is 230.
It indicates the possibility that
  response-biased sampling can be designed  more efficiently than  prospective sampling
  in terms of parameter estimation.

Overall, the results of simulation studies agree with the theory. Moreover,
the consistency and asymptotic normality established in Theorem 1 might hold in more
general scenarios, without the  technical conditions.
\par

\setcounter{chapter}{5}
\setcounter{equation}{0} %-1
\noindent {\bf 5. Applications}

In this section, we apply the proposed method to the Forbes Global 2000 data  published in 2012 and the Stanford heart transplant data.
%\begin{center}
%\textsc{Forbes Global 2000 data}
%\end{center}
%In data set, \textit{sales}, \textit{profits}, \textit{assets} and \textit{market values} of largest 2000 companies worldwide are listed.

The first data set contains the profits, assets and market value for companies
of the Forbes Global 2000.
It is commonly known that
profits is a measure of the financial performance of the companies and assets indicate the size of the companies.
 The purpose of this study is to analyze the relationship among market value, profits and assets
of companies with the existing Forbes Global 2000 data. But the companies on the Forbes Global 2000 list,
 that are the biggest and most powerful companies in the world,
is in fact a biased sampling data from the population.  The sample size here $n=2000$.
 We fit the transformation model to the data with
covariates  $X_1 =$ \textit{assets}$/250$, $Z =$ \textit{profits} and response
$Y$=\textit{market value} with the proposed method. For identifiability, we set the coefficient of $Z$ to $1$.
The random weights are generated from the standard
exponential distribution with resampling times $N=500$. The estimate
of the coefficient of $X_1$ is 0.2912 and the estimated standard error is
0.0503.

Our second example pertains to the Stanford heart transplant data.
Crowley and Hu (1977) reported information of 103 potential heart transplant recipients in the Stanford heart transplantation program consisting age, waiting time to transplantation, survival or censoring time from acceptance to the programme, and three mismatch scores  from October 1967 to April 1974. During that time, 69 of the patients underwent the operation. Miller and Halpern (1982) reported the survival times, censored or uncensored in February 1980 of 184 patients who had received heart transplants.
%27 patients without mismatch scores and 5 patients with survival times less than 10 days were deleted from the data in Miller and Halpern (1982).
%We can treat the data as left-truncated and right-censored if we only analyse the data of transplant patients.
%And if we choose the 97 observations with their survival time uncensored, the data set becomes response-biased.
Similar to Miller and Halpern (1982),
we consider the 152 patients whose $T5$ mismatch scores were not missing and survival times were not less than 10 days.
Thus the observations in the study
 are  left-truncated and right-censored. In view of remark 3 in section 3, the 97 complete observations can be treated as drawn
 from a response-biased sampling.
 %We regress the survival times against the ages, age$^2$s and $T5$ mismatch scores. By the insignificance of the $T5$ mismatch score, we fix its coefficient to be $10^{-6}$. Our proposed method gives the estimating result of $(\hat\beta_{age}, \hat{\beta}_{age^2})$ being $(-0.2220,0.0018)$ with standard error $(0.0238,0.0011)$. We conclude that
   %the survival time is negatively related with the age.
 We regress the \textit{survival time} against \textit{age} and \textit{age$^2$} with the transformation model. The coefficient of \textit{age} is fixed to $1$ and
 the random weights are generated from standard
exponential distribution with 500 replications.
Our proposed method gives the estimate of the coefficient of \textit{age$^2$} being $-0.0152$ with standard error $0.0031$ and the $95\%$ confidence interval $[-0.0213,-0.0091]$.
%Therefore we conclude that the positive effect of \textit{age$^2$} on \textit{survival time} is significant.
To compare with the Cox's estimator presented in Miller and Halpern (1982), we consider the ratio of the coefficient of \textit{age$^2$} to the coefficient of \textit{age}. The resulting estimate is $-0.0161$ with standard error $0.0016$ and the $95\%$ confidence interval $[-0.0191,-0.0130]$.
%Therefore we conclude that the positive effect of \textit{age$^2$} on \textit{survival time} is significant.
%Both methods conclude that the negative effect of \textit{age$^2$} on \textit{survival time} is significant.
It can be seen from both methods that the confidence interval, which does not cover zero, confirms the negative quadratic effect
of the \textit{age}.
Note that the confidence interval of Cox's estimator is contained inside that obtained from the proposed method. This is mainly because the sample size from response-biased sampling is smaller and the transformation model is more general than the Cox's model. In addition, a comparison among different methods applying to an earlier published Standard heart transplant data set can be found in Khan and Tamer (2007), which also gives a similar estimating result to the MRC method.
\par

\setcounter{chapter}{6}
\setcounter{equation}{0} %-1
\noindent {\bf 6. Concluding remarks}

This paper gives a general method of regression analysis based on
the method of MRC for transformation models with response-biased sampling. Consistency and asymptotic normality of the proposed estimator are proved theoretically. Simulation studies show that response-biased sampling gives a more efficient estimation than prospective sampling in certain situations, and the proposed estimator works well for
 a variety of sampling schemes and models. In addition, the nature of the MRC method implies
 that the estimation does not vary with different monotonic transformations, avoiding
   the estimation of the transformation functions.
%\item
%The condition $\frac{\partial^2}{\partial t^2}(\log{f_{\epsilon^*}(t)})<0 $ for all
%$t$ may seem to be quite strong, but it includes $N(0,\sigma^2)$ and Pareto family, Cox model and
%proportional odds model. Furthermore, we also tested several error distributions which violate this condition in the simulation and their performances are rather good. So we are confident that this condition can be loosen a little bit.
%\item Assumption of independence of $Z^*$ and $X^*$ may seem to be restrictive. It can be largely overcomed
%by first orthogonal decomposition of the covariate $(Z,X)$.
Furthermore, this method can be applied to more general  models of the form
 \begin{eqnarray}
\label{1-2} Y^*=D \cdot F(\theta'_0 W^*,\epsilon^*),
\end{eqnarray}
where $Y^*$, $W^*$, $\theta_0$ and $\epsilon^*$ are defined as in section 2. $D\colon \mathbb{R} \rightarrow \mathbb{R}$ is a non-degenerate, monotonic function and $F \colon \mathbb{R}^2 \rightarrow \mathbb{R}$ is strictly monotonic in each of the variables. Though we cannot separate the covariate term and the error term in this model, our estimation and inference procedure can still be applied as long as the monotonicity assumptions of the composite transformation $D \cdot F$ are valid.
\par
%%%%%%%%%%%%%%%%%%%%%%%%%%%%%%%%%%%%%%%%%%%%%%%%%%%%%%%%%%%%%%%%%%%%%%%%%%%%%%%%%%%%%%%%%%%%%%%%%%%%%%%%%%%%%%%%%%%%%%%%%%%%

%\noindent {\large\bf Acknowledgment}
%Write the acknowledgment here.
%\par

%%%%%%%%%%%%%%%%%%%%%%%%%%%%%%%%%%%%%%%%%%%%%%%%%%%%%%%%%%%%%%%%%%%%%%%%%%%%%%%%%%%%%%%%%%%%%%%%%%%%%%%%%%%%%%%%%%%%%%%%%%%%

\noindent{\large\bf References}
\begin{description}
\item  {\sc Abrevaya, J.}
(1999). Computation of the maximum rank correlation estimator. {\it Economics letters}  {\bf 62}, 279--285.
\item  {\sc Anderson, J. A.}
(1972). Separate sample logistic discrimination. {\it Biometrika}  {\bf 59}, 19--35.
\item {\sc Bickel, P. J. \& Ritov, Y.} (1991).
Large sample theory of estimation in biased sampling regression models.
{\it Ann. Statist.}  {\bf 19}, 797--816.
\item {\sc Binder, D. A.} (1992). Fitting Cox's proportional hazards models from survey data. {\it Biometrika}  {\bf
79}, 139--147.
\item {\sc Breslow, N. E. \& Day, N. E.} (1980).
{\it The Analysis of Case-Control Studies.}
Lyon: International Agency for Research on Cancer.

\item {\sc Chen, K.} (2001).
Parametric models for response-biased sampling.
{\it J. R. Statist. Soc. B.} {\bf 63}, 775--789.

\item {\sc Chen, K., Jin, Z.  \&  Ying, Z.} (2002). Semiparametric analysis
of transformation model with censored data. {\it Biometrika} {\bf
89}, 659--668.

\item {\sc  Cheng, S. C., Wei, L. J.   \&   Ying, Z.} (1995).
Analysis of transformation models with censored data. {\it
Biometrika} {\bf 82}, 835--845.

\item  {\sc  Cheng, S. C., Wei, L. J.   \&   Ying, Z.} (1997).
Prediction of survival probabilities with semi-parametric transformation models. {\it
J. Am. Statist. Assoc.} {\bf 92}, 227--235.

\item {\sc  Cosslet, S.R.} (1981). Maximum likelihood estimate for choice-based samples. {\it Econometrica}  {\bf 49}, 1289--1316.

%\item {\sc  Crowley, J. \& Hu, M.} (1977). Covariance analysis of heart transplant survival data. {\it J. Am. Statist. Assoc.}  {\bf 49}, 1289--1316.

\item {\sc Dabrowska, D. M.  \&  Doksum, K. A.} (1988).
Estimation and testing in the two-sample generalized odds-rate model. {\it  J. Am. Statist. Assoc.} {\bf 83}, 744--749.

\item {\sc Han, A. K.} (1987).
Non-parametric analysis of a generalized regression model. {\it  J. Econometrics} {\bf 35}, 303--316.

\item {\sc Hausman, J. A. \&   Wise, D. A.} (1981).
Stratification on endogenous variables and estimation: the Gary Income Maintenance Experiment. In {\it Structural Analysis of Discrete Data: with Econometric Applications} (eds C. Manski and D. McFadden), pp. 364--391. Cambridge: Massachusetts Institute of Technology Press.

\item {\sc Heckman, J. J. } (1977). Sample selection bias as a specification error with an application to the estimation of labor supply functions. NBER working paper \#172.

\item {\sc Heckman, J. J. } (1979). Sample selection bias as a specification error. {\it Econometrica} {\bf 47}, 153--161.

\item {\sc Horvitz, D. G.  \&  Thompson, D. J.} (1952).
A generalization of sampling without replacement from a finite universe. {\it  J. Am. Statist. Assoc.} {\bf 47}, 663--685.
%\item {\sc Jewell, N.} (1985).
%{\sl Information Bounds and Nonparametric
%Maximum Likelihood Estimation.}  Basel: Birkh$\ddot{a}$user.

\item {\sc Huang, C. Y. \& Qin, J. } (2010). Nonparametric estimation for length-biased and right-censored data. {\it Biometrika} {\bf 98},
177--186.

\item {\sc Jewell, N.} (1985). Regression from stratified samples of dependent variables. {\it Biometrika} {\bf 72}, 11--21.

\item {\sc Jin, Z., Ying, Z. \& Wei, L. J.} (2001).
A simple resampling method by perturbing the minimand. {\it Biometrika} {\bf  88}, 381--390.

\item {\sc Khan, S. \& Tamer, E.} (2007).
Partial rank estimation of duration models with general forms of censoring. {\it J. Econometrics} {\bf  136}, 251--280.

\item {\sc Lawless, J. F., Kalbfleisch, J. D. \&
Wild, C. J. } (1999). Semiparametric methods for response-selective
and missing data problems in regression. {\it J. R. Statist.
Soc. B.} {\bf 61}, 413--438.

\item {\sc Lin, D. Y.} (2000). On fitting Cox's proportional hazards models to survey data. {\it Biometrika} {\bf
87}, 37--47.

\item {\sc Luo, X., \& Tsai, W. Y.} (2009). Nonparametric estimation for right-censored length-biased data: a pseudo-partial likelihood approach.
{\it Biometrika} {\bf 96}, 873--886.

\item {\sc Luo, X., Tsai, W. Y. \& Xu, Q.} (2009). Pseudo-partial likelihood estimators for the Cox regression model with missing covariates. {\it Biometrika} {\bf 96}, 617--633.

\item {\sc Manski, C. F.} (1993).
The selection problem in econometrics and statistics. \\{\it Econometrika} (eds G. S. Maddala, C. R, Rao and H. D. Vinod), pp. 73--84. Amsterdam: North-Holland.

\item {\sc Manski, C. F.   \&  Lerman, S.} (1977). The estimation of choice probabilities from choice-based samples. {\it Econometrica} {\bf 45}, 1977--1988.

\item {\sc Miller, R.   \&  Halpern, J.} (1982). Regression with censored data. {\it Biometrika} {\bf 69}, 521--531.

\item {\sc Nelder, J. A. \& Mead, R.} (1965). A simplex algorithm for function minimization. {\it Computer Journal} {\bf 7}, 308-313.

\item {\sc Ning, J., Qin, J. \& Shen, Y. } (2010). Nonparametric tests for right-censored
data with biased sampling. {\it J. R. Statist. Soc. B.}  {\bf 72}, 609-630.

\item {\sc Prentice, R. L. \& Pyke, R.} (1979). Logistic disease incidence models with case-control studies. {\it Biometrika} {\bf 66}, 403--411.

\item {\sc Scott, A. J. \& Wild, C. J.} (1986). Fitting logistic models under case-control or choice based sampling. {\it J. R. Statist. Soc. B.} {\bf 48}, 170--182.

\item {\sc Sherman, R.} (1993).
The limiting distribution of the maximum rank correlation estimator.
{\it Econometrica}  {\bf 61}, 123--137.

\item {\sc Sherman, R.} (1994).
Maximal Inequalities for Degenerate U-processes with Applications to Optimization Estimators.
{\it Ann. Statist.} {\bf 22}, 439--459.

\item {\sc Shen, Y., Ning, J. \& Qin, J. } (2009). Analyzing length-biased data with semiparametric transformation and accelerated failure time models.
{\it  J. Am. Statist. Assoc.} {\bf 104}, 1192--1202.

\item {\sc Song, X., Ma, S., Huang, J. \& Zhou, X.H. } (2007). A semiparametric approach for the nonparametric transformation survival model with multiple covariates. {\it Biostatistics}
{\bf 8}, 197--211.

\item {\sc Stone, C. J. } (1980). Optimal rates of convergence for
nonparametric estimators. {\it Ann. Statist.}  {\bf 8}, 1348--1360.

\item {\sc Stone, C. J.} (1982). Optimal global rates of convergence for
nonparametric regression. {\it Ann. Statist.}  {\bf 10}, 1040--1053.

\item {\sc Tsai, W. Y.} (2009). Pseudo-partial likelihood for proportional hazards models with biased-sampling data.  {\it Biometrika} {\bf 96}, 601--615.

%\item {\sc Wang, M. C.}(1996). Hazards regression analysis for length-biased data. {\it Biometrika} {\bf 83}, 343--354.

\item {\sc Van der Vaart, A.  \&   Wellner, J. A.}
 (1996). {\sl  Weak Convergence and Empirical
Processes: With Applications to Statistics.} Springer-Verlag.

\item {\sc Wang, M. C.}(1996). Hazards regression analysis for length-biased data. {\it Biometrika} {\bf 83}, 343--354.

\item {\sc Zeng, D. \& Lin, D. Y. }(2007). Maximum likelihood estimation in semiparametric
regression models with censored data. {\it J. R. Statist. Soc. B.} {\bf 69}, 507--564.
\end{description}

%%%%%%%%%%%%%%%%%%%%%%%%%%%%%%%%%%%%%%%%%%%%%%%%%%%%%%%%%%%%%%%%%%%%%%%%%%%%%%%%%%%%%%%%%%%%%%%%%%%%%%%%%%%%%%%%%%%%%%%%%%%%

\vskip .65cm
\noindent
Kani Chen, Department of Mathematics, Hong Kong University of Science
and Technology, Hong Kong
\vskip 2pt
\noindent
E-mail: makchen@ust.hk
\vskip 2pt

\noindent
Yuanyuan Lin, Department of Applied Mathematics, Hong Kong Polytechnic University, Hong Kong
\vskip 2pt
\noindent
E-mail: yy.lin@polyu.edu.hk

\noindent
Yuan Yao, Department of Mathematics, Hong Kong Baptist University, Hong Kong
\vskip 2pt
\noindent
E-mail: yaoyuan@hkbu.edu.hk

\noindent
Chaoxu Zhou, Department of Mathematics, Hong Kong University of Science
and Technology, Hong Kong
\vskip 2pt
\noindent
E-mail: chaoxu.zhou@gmail.com
\vskip .3cm

%%%%%%%%%%%%%%%%%%%%%%%%%%%%%%%%%%%%%%%%%%%%%%%%%%%%%%%%%%%%%%%%%%%%%%%%%%%%%%%%%%%%%%%%%%%%%%%%%%%%%%%%%%%%%%%%%%%%%%%%%%%%
%%%%%%%%%%%%%%%%%%%%%%%%%%%%%%%%%%%%%%%%%%%%%%%%%%%%%%%%%%%%%%%%%%%%%%%%%%%%%%%%%%%%%%%%%%%%%%%%%%%%%%%%%%%%%%%%%%%%%%%%%%%%
\newpage
%\begin{center}
%\textsf{ \hspace{0.1in} APPENDIX: Proof of the Theorem }
%\end{center}
\setcounter{chapter}{7}
\setcounter{equation}{0}
\noindent {\bf Appendix: Proof of Theorem 1}
\vspace{0.1in}

Consider the transformation model
\begin{eqnarray}
 \label{A-1}H(Y^*)=\theta'_0W^* +\epsilon^*,
\end{eqnarray}
where $H(\cdot)$ is an unknown monotonically increasing   function,   $\epsilon^*$ is the error,
 independent of $W^*$, with unspecified distribution, and
$\theta_0$ is a $(d+1)$-dimensional vector of regression coefficients.
Accordingly, $W^*$ can be decomposed into $W=(Z^*,X^*)$, where $Z^*$ is the covariate corresponding to the fixed regression coefficient and $X^*$ is the other $d$-dimensional covariate.
Hence, the model can be rewritten as
 \begin{eqnarray}
H(Y^*)=Z^*+\beta'_0X^* +\epsilon^*\nonumber.
\end{eqnarray}

We point out that the parameter estimation does not vary with different decompositions of covariates. Suppose that $(\tilde Z^*, \tilde X^*)$ is another composition of covariates. Then there exists a unique matrix $P$ of full rank such that
$(\tilde Z^*, \tilde {X^*}')'=P(Z^*,{X^*}')'$.

Suppose that
$\tilde\beta'(\tilde Z^*, \tilde {X^*}')'=\beta'(Z^*,{X^*}')'$. Then by the uniqueness of linear representation, the relevant parameter must satisfy that $P\tilde\beta =\beta$. So if one parameter is uniquely determined in a $d$-dimensional linear space, the other parameter is also uniquely determined in a transformed $d$-dimensional linear space.

For easier explanation in the technical proof, we rewrite the transformation model (\ref{A-1}) into
$$H(Y^*)=Z^*+\beta_0' X^*+\epsilon^*,$$
where we suppose the covariance decomposition satisfies that $\tilde Z^*:=Z^*+\beta_0' X^*$ is irrelevant of $X^*$. Such a decomposition always exists since $\theta'_0W^*$ is a one-dimensional vector in a $(d+1)$-dimensional linear space, so it has a $d$-dimensional orthogonal compliment which can be defined as $X^*$. Furthermore, $\tilde Z^*$ and $X^*$ are supposed to be independent.

Suppose the regularity conditions hold:\\
C1) The unknown parameter $\beta$ lies in a bounded space $\mathbf{B}\subset \mathcal {R}^d$;\\
C2) Both of $Z^*$ and $X^*$ have continuously differentiable density functions to the second order;\\
C3) $f_{\epsilon^*}$ is log-concave (i.e., $\log f_{\epsilon^*}$ is concave);\\
C4) (Identifiability condition) $\xi(\beta):=(\beta-\beta_0)'(X_2^*-X_1^*)=0$ almost surely if and only if $\beta=\beta_0$.

\noindent \textit{Consistency:}

Define $$g(\beta)=E[ I\{Y_1<Y_2\}I\{\beta X_1+Z_1<\beta X_2+ Z_2\}]$$ and
$$g_n(\beta)=\frac{1}{n^2-n}\sum_{i\neq j}I\{Y_i<Y_j\}I\{\beta X_i+Z_i<\beta X_j+ Z_j\}.$$

\noindent Step 1. We show that $g(\beta)$ has a unique maximum at $\beta=\beta_0$.

Write, for any $t_1<t_2$,
\begin{eqnarray}
& &E[I\{Y_1<Y_2\}I\{\beta X_1+ Z_1<\beta X_2+ Z_2\}|Y_1=t_1,Y_2=t_2]\nonumber\\
&=&P(\beta X_1+ Z_1<\beta X_2+ Z_2|Y_1=t_1,Y_2=t_2)\nonumber\\
&=&P(\beta X_1^*+ Z_1^*<\beta X_2^*+ Z_2^*|Y_1^*=t_1,Y_2^*=t_2)\nonumber\\
&=&P(Z_1^*-Z_2^*<\beta X_2^*-\beta X_1^*|\beta_0 X_1^*+ Z_1^*+\epsilon_1^*=H(t_1),\beta_0 X_2^*+ Z_2^*+\epsilon_2^*=H(t_2))\nonumber\\
%&=&P(X_1+\beta_0 Z_1-(X_2+\beta_0 Z_2)|X_1+\beta_0 Z_1+\epsilon_1=H(t_1),X_2+\beta_0 Z_2+\epsilon_2=H(t_2))\nonumber\\
&=&P(\tilde Z_1^*-\tilde Z_2^*<(\beta-\beta_0)(X_2^*-X_1^*)|\tilde Z_1^*+\epsilon_1^*=\tilde t_1,\tilde Z_2^*+\epsilon_2^*=\tilde t_2)\nonumber\\
&=&\frac{\int P(\xi(\beta)>s_1-s_2)f_{\tilde Z^*} (s_1)f_{\epsilon^*}(\tilde t_1-s_1)f_{\tilde Z^*} (s_2)f_{\epsilon^*}(\tilde t_2-s_2)d s_1 d s_2}{\int f_{\tilde Z^*} (s)f_{\epsilon^*}(\tilde t_1-s)d s\int f_{\tilde Z^*} (s)f_{\epsilon^*}(\tilde t_2-s)d s},
\end{eqnarray}
where $\tilde t_i=H(t_i)$, $i=1,2$.

The denominator is irrelevant with $\beta$. The numerator will be proved to have a unique maximum at $\beta=\beta_0$. The numerator can be written as
\begin{eqnarray}
& &\frac{1}{2}\int [1-sgn(s_1-s_2)P(|\xi(\beta)|<|s_1-s_2|)]\nonumber\\
& & \hskip 6cm f_{\tilde Z^*}(s_1)f_{\epsilon^*}(\tilde t_1-s_1)f_{\tilde Z^*}(s_2)f_{\epsilon^*}(\tilde t_2-s_2)d s_1 d s_2\nonumber\\
&=&\frac{1}{2}\int f_{\tilde Z^*}(s_1)f_{\epsilon^*}(\tilde t_1-s_1)f_{\tilde Z^*}(s_2)f_{\epsilon^*}(\tilde t_2-s_2)d s_1 d s_2+\Pi(\beta)\nonumber
\end{eqnarray}
where
\begin{eqnarray}
\Pi(\beta)&=&-\frac{1}{2}\int sgn(s_1-s_2)P(|\xi(\beta)|<|s_1-s_2|)\nonumber\\
&& \hskip 5cm f_{\tilde Z^*}(s_1)f_{\epsilon^*}(\tilde t_1-s_1)f_{\tilde Z^*}(s_2)f_{\epsilon^*}(\tilde t_2-s_2)d s_1 d s_2\nonumber.
\end{eqnarray}

It then suffices to show that $\Pi(\beta)$ is uniquely maximized at $\beta=\beta_0$. To this end, write
\begin{eqnarray}
\Pi(\beta)&=&\frac{1}{2}\int_{s_1<s_2}g_\beta^*(|s_1-s_2|)f_{\tilde Z^*}(s_1)f_{\epsilon^*}(\tilde t_1-s_1)f_{\tilde Z^*}(s_2)f_{\epsilon^*}(\tilde t_2-s_2)d s_1 d s_2\nonumber\\
& &-\frac{1}{2}\int_{s_1>s_2}g_\beta^*(|s_1-s_2|)f_{\tilde Z^*}(s_1)f_{\epsilon^*}(\tilde t_1-s_1)f_{\tilde Z^*}(s_2)f_{\epsilon^*}(\tilde t_2-s_2)d s_1 d s_2\nonumber\\
&=&\frac{1}{2}\int_{s_1<s_2}g_\beta^*(|s_1-s_2|)f_{\tilde Z}(s_1)f_{\tilde Z}(s_2)\nonumber\\
&& \hskip 3cm [f_{\epsilon^*}(\tilde t_1-s_1)f_{\epsilon^*}(\tilde t_2-s_2)-f_{\epsilon^*}(\tilde t_1-s_2)f_{\epsilon^*}(\tilde t_2-s_1)]d s_1 d s_2,\nonumber\\
\end{eqnarray}
where we define $g_\beta^*(t)= P(|\xi(\beta)|<
t)$ and then $g_{\beta_0}^*= 1$ since $\xi(\beta_0)\equiv 0$.

Since $g^*(\cdot)$ is only maximized at $\beta=\beta_0$ by assumption, to show that $\beta_0$ is the unique maximizer of $g(\beta)$, we only need to prove that the quantity in the square brackets is positive for all $\tilde t_1<\tilde t_2$ and $s_1<s_2$.

Now we show
$$h(\tilde t_1-s_1)+h(\tilde t_2-s_2)>h(\tilde t_1-s_2)+h(\tilde t_2-s_1)$$
for all $\tilde t_1<\tilde t_2$ and $s_1<s_2$, where $h=\log f_\epsilon$.

By the fact that $f_{\epsilon^*}$ is log-concave,
$$\frac{\partial}{\partial t}(h(t-s_1)-h(t-s_2))=\int_{t-s_2}^{t-s_1}\frac{d^2}{d s^2}h(s)d s<0.$$
Therefore $h(t-s_1)-h(t-s_2)$ is decreasing in $t$.
As a result, $$h(\tilde t_1-s_1)+h(\tilde t_2-s_2)>h(\tilde t_1-s_2)+h(\tilde t_2-s_1).$$

\noindent Step 2. We show that
\begin{eqnarray}
\label{A-10}
\sup_\beta|g_n(\beta)-g(\beta)|=O_p(\sqrt{\frac{\log n}{n}}).
\end{eqnarray}

For each $n\in \mathcal{N}$,
let $\{\beta_{n_1},\cdots, \beta_{n_m}\}$ be a $1/n^2$-net of $\mathbf{B}$, which means that $$\mathbf{B}\subset \cup_{k=1}^m B (\beta_{n_k},\frac{1}{n^2}).$$
Then $m=O(n^{2d})$.

For $M>1$, we have
\begin{eqnarray}
\label{A-11}
& &P(\sup_\beta[g_n(\beta)-g(\beta)]>M\sqrt{\frac{\log n}{n}})\nonumber\\
&\leq&P(\sup_{k=1,\cdots, m}[g_n(\beta_{n_k})-g(\beta_{n_k})]>(M-1)\sqrt{\frac{\log n}{n}})\nonumber\\
& &+P(\sup_\beta[g_n(\beta)-g(\beta)]-\sup_{k=1,\cdots, m}[g_n(\beta_{n_k})-g(\beta_{n_k})]>\sqrt{\frac{\log n}{n}}).
\end{eqnarray}

By Hoeffding's inequality (1963) for U-statistics, the first term in the right hand side of (\ref{A-11}) can be bounded by $O(n^{2d-(M-1)^2/4})$. Using Chebyshev's inequality, the second term  in the right hand side of (\ref{A-11}) is bounded by $O(\frac{1}{n^2})$.

Now we have shown that
\begin{eqnarray}
& &P(\sup_\beta[g_n(\beta)-g(\beta)]>M\sqrt{\frac{\log n}{n}})\nonumber\\
&=&O(n^{2d-(M-1)^2/4})+O(\frac{1}{n\log n}).
\end{eqnarray}

Since the last equality still holds if we replace $g_n$ and $g$ by $-g_n$ and $-g$,
it can be written as
\begin{eqnarray}
& &P(\sup_\beta|g_n(\beta)-g(\beta)|>M\sqrt{\frac{\log n}{n}})\nonumber\\
&=&O(n^{2d-(M-1)^2/4})+O(\frac{1}{n\log n}).
\end{eqnarray}

Then it follows equality (\ref{A-10}).

\noindent Step 3. We show that $\hat\beta_n$ converges to $\beta_0$ in probability.

Since $\beta_0$ is the unique maximizer of $g$, and $\hat\beta_n$ is the maximizer of $g_n$, we have
\begin{eqnarray}
0&\leq&g(\beta_0)-g(\hat\beta_n)\nonumber\\
&=&[g(\beta_0)-g_n(\beta_0)]-[g(\hat\beta_n)-g_n(\hat\beta_n)]-
[g_n(\hat\beta_n)-g_n(\beta_0)]\nonumber\\
&\leq&[g(\beta_0)-g_n(\beta_0)]-[g(\hat\beta_n)-g_n(\hat\beta_n)]\nonumber\\
&=&O_p(\sqrt{\frac{\log n}{n}})+O_p(\sqrt{\frac{\log n}{n}})\nonumber\\
&=&O_p(\sqrt{\frac{\log n}{n}})
\end{eqnarray}

 On the other hand, by the differentiability of density functions of $\tilde Z$ and $X$, note that $\beta_0$ is the unique maximizer of $g$ and $\dot g(\beta_0)=0$, the Taylor expansion can then be written as
\begin{eqnarray}
g(\hat\beta_n)-g(\beta_0)=-(\hat\beta_n-\beta_0)'A(\hat\beta_n-\beta_0)+o_p(\hat\beta_n-\beta_0)^2,
\end{eqnarray}
where $A$ is a positive definite matrix.

Compare the last two equations, it follows that
\begin{eqnarray}
\hat\beta_n-\beta_0=O_p(\sqrt[4]{\frac{\log n}{n}})=o_p(n^{-1/5}).
\end{eqnarray}

The consistency is proved.

\noindent \textit{Asymptotic normality:}

We still use the notation of $g$ and $g_n$ as above. Furthermore, denote
\begin{eqnarray}
\epsilon_n(\beta)=g_n(\beta)-g(\beta).
\end{eqnarray}

Standard decomposition of U-statistics gives
\begin{eqnarray}
\epsilon_n(\beta)-\epsilon_n(\beta_0)=\frac{1}{n}\sum_{i=1}^n b_i(\beta)+\frac{1}{n^2-n}\sum_{i<j}d_{ij}(\beta),
\end{eqnarray}
where
\begin{eqnarray}
b_i(\beta)=E[a_{ij}(\beta)+a_{ji}(\beta)
-2E a_{ij}(\beta)|Z_i,X_i,Y_i],
\end{eqnarray}

\begin{eqnarray}
d_{ij}(\beta)=a_{ij}(\beta)+a_{ji}(\beta)
-2E a_{ij}(\beta)-b_i(\beta)-b_j(\beta).
\end{eqnarray}

and
\begin{eqnarray}
& &
a_{ij}(\beta)=[I\{Z_i+\beta'X_i>Z_j+\beta'X_j\}-I\{Z_i+\beta_0'X_i>Z_j+
\beta_0'X_j\}]
\nonumber\\
& &\hskip 9cm
I\{Y_i>Y_j\}.
\end{eqnarray}

Note that $E b_i(\beta)\equiv0$,  Taylor expansion gives
\begin{eqnarray}
\frac{1}{n}\sum_{i=1}^n b_i(\beta)=(\beta-\beta_0)'\frac{1}{n}\sum_{i=1}^n \dot{b}_i(
\beta_0)+o_p(|\beta-\beta_0|)^2.
\end{eqnarray}

Using exponential inequality again, similar to the step 2 in the proof of consistency, we have
\begin{eqnarray}
\sup_{|\beta-\beta_0|=o_p(n^{-1/5})}|\frac{1}{n^2-n}\sum_{i<j}d_{ij}(\beta)|
=o_p(n^{-1}).
\end{eqnarray}

So far we have shown that
\begin{eqnarray}
& &g_n(\beta)\nonumber\\&=&g(\beta)+\epsilon_n(\beta)\nonumber\\
&=&g(
\beta_0)-\frac{1}{2}(\beta-\beta_0)'A(\beta-\beta_0)
+(\beta-\beta_0)'\frac{1}{n}\sum_{i=1}^n
\dot b_i(\beta_0)+\epsilon_n(\beta_0)
+o_p(|\beta-\beta_0|)^2\nonumber\\
& &\hskip 11cm+o_p(n^{-1})\nonumber\\
&=&f_n(\beta)+\epsilon_n(\beta_0)
+o_p(n^{-1}),
\end{eqnarray}
where
\begin{eqnarray}
& &f_n(\beta)\nonumber\\
&=&g(\beta_0)-\frac{1}{2}(\beta-\beta_0)'A(\beta-\beta_0)
+(\beta-\beta_0)'\frac{1}{n}\sum_{i=1}^n
\dot b_i(\beta_0)+o_p(|\beta-\beta_0|)^2\nonumber\\
&=&g(\beta_0)-\frac{1}{2}(\beta-\beta_0)'A_n(\beta-\beta_0)
+(\beta-\beta_0)'\frac{1}{n}\sum_{i=1}^n
\dot b_i(\beta_0)\nonumber\\
&=&g(\beta_0)-\frac{1}{2}\{A_n^{1/2}[\beta-\beta_0-A_n^{-1}\frac{1}{n}
\sum_{i=1}^n\dot b_i(\beta_0)]\}'\{A_n^{1/2}[\beta-\beta_0-A_n^{-1}\frac{1}{n}
\sum_{i=1}^n\dot b_i(\beta_0)]\}
\nonumber\\
& &+\frac{1}{2}(\frac{1}{n}\sum_{i=1}^n\dot b_i(\beta_0))'A_n^{-1}(\frac{1}{n}\sum_{i=1}^n\dot b_i(\beta_0)),
\end{eqnarray}
where we let $o_p(|\beta-\beta_0|)^2=c_n|\beta-\beta_0|^2$ with $c_n=o_p(1)$ and $A_n=A-2c_n I$.

So the maximizer of $f_n$ is
\begin{eqnarray}
\hat\gamma_n=\beta_0+A_n^{-1}\frac{1}{n}\sum_{i=1}^n\dot b_i(\beta_0)
%=\beta_0+O_p(n^{-1/2}),
\end{eqnarray}

Suppose that $\hat\beta_n$ is the maximizer of $g_n$, then
\begin{eqnarray}
\label{A-27}
0&\leq&f_n(\hat\gamma_n)-f_n(\hat\beta_n)\nonumber\\
&=&[f_n(\hat\gamma_n)+\epsilon_n(\beta_0)-g_n(\hat\gamma_n)]
-[f_n(\hat\beta_n)+\epsilon_n(\beta_0)-g_n(\hat\beta_n)]-[g_n(\hat\beta_n)
-g_n(\hat\gamma_n)]\nonumber\\
&\leq&[f_n(\hat\gamma_n)+\epsilon_n(\beta_0)-g_n(\hat\gamma_n)]
-[f_n(\hat\beta_n)+\epsilon_n(\beta_0)-g_n(\hat\beta_n)]\nonumber\\
&=&o_p(n^{-1})+o_p(n^{-1})\nonumber\\
&=&o_p(n^{-1}).
\end{eqnarray}

On the other hand, from the expression of $f_n$,
\begin{eqnarray}
\label{A-28}
& &f_n(\hat\gamma_n)-f_n(\hat\beta_n)\nonumber\\
&=&\frac{1}{2}\{A_n^{1/2}[\hat\beta_n
-\beta_0-A_n^{-1}\frac{1}{n}
\sum_{i=1}^n\dot b_i(\beta_0)]\}'\{A_n^{1/2}[\hat\beta_n-\beta_0-A_n^{-1}\frac{1}{n}
\sum_{i=1}^n\dot b_i(\beta_0)]\}.\nonumber\\
\end{eqnarray}

Compare (\ref{A-27}) and (\ref{A-28}), finally we have
\begin{eqnarray}
\hat\beta_n&=&\beta_0+A_n^{-1}\frac{1}{n}\sum_{i=1}^n \dot b_i(\beta_0)
+o_p(n^{-1/2})\nonumber\\
&=&\beta_0+A^{-1}\frac{1}{n}\sum_{i=1}^n \dot b_i(\beta_0)+(A_n^{-1}-A^{-1})\frac{1}{n}\sum_{i=1}^n \dot b_i(\beta_0)
+o_p(n^{-1/2})\nonumber\\
&=&\beta_0+A^{-1}\frac{1}{n}\sum_{i=1}^n \dot b_i(\beta_0)+o_p(n^{-1/2}),
\end{eqnarray}
where the last equation comes from that $$A_n^{-1}-A^{-1}=o_p(1)$$ and $$\frac{1}{n}\sum_{i=1}^n \dot b_i(\beta_0)=O_p(n^{-1/2}) $$ by the definition of $A_n$ and the central limit theorem.

Therefore,
$$\sqrt{n}(\hat\beta_n-\beta_0)=A^{-1}\frac{1}
{\sqrt{n}}\sum_{i=1}^n \dot b_i(\beta_0)+o_p(1)\rightarrow N(0, \Sigma)$$ in distribution, where
$$\Sigma=A^{-1}Var\{\dot b_1(\beta_0)\}(A^{-1})'.$$
\vspace{2 in}

\begin{table}
{\scriptsize
%\begin{center}
%\textbf{Table 1:} \textit{Simulation results for the proposed method
%with different models and sampling schemes. }
\caption{\textit{Simulation results for different models and sampling schemes.}}
\begin{tabular*}{\textwidth}{c|ccccccc|cccccc}
\cline{1-14}
\cline{1-14}
S& \multicolumn{5}{c}{Proposed}&\multicolumn{2}{c|}{IP}& \multicolumn{4}{c}{Proposed}&\multicolumn{2}{c}{IP}\\
\cline{1-14}

 & & BIAS & SE & SEE & CP & BIAS & SE &  BIAS & SE & SEE & CP &BIAS & SE\\
\cline{2-14}
 & & \multicolumn{5}{c}{$\epsilon\sim$ Double Exponential}&&
  \multicolumn{6}{c}{$\epsilon\sim$ Extreme Value} \\

& $\beta_1$ & 0.0238 & 0.2213 & 0.2286 & 0.97 &0.3004&0.0966& 0.0390& 0.1443 & 0.1632& 0.98&-0.2433&0.1031 \\
1 & $\beta_2$ & 0.0030 & 0.2337 & 0.2320 & 0.99 &-0.3617&0.1459& 0.0234& 0.1339 & 0.1638& 0.97& -0.2039 & 0.1438 \\
\cline{2-14}
& & \multicolumn{5}{c}{$\epsilon\sim$ Normal }&&
  \multicolumn{6}{c}{ $\epsilon\sim$ Logistic} \\
& $\beta_1$ & 0.0159 & 0.1338 & 0.1368 & 0.98 &0.2149&0.0540& 0.0084& 0.1933 & 0.1826& 0.94 & 0.5993& 0.0918\\
& $\beta_2$ & 0.0234 & 0.1430 & 0.1356 & 0.97 &-0.2661 & 0.0603& 0.0041& 0.2117 & 0.1901& 0.92 &-0.3347 & 0.1376 \\
\cline{1-14}

 & & \multicolumn{5}{c}{$\epsilon\sim$ Double Exponential}&&
  \multicolumn{6}{c}{$\epsilon\sim$ Extreme Value} \\
& $\beta_1$ & 0.0142 & 0.1887 & 0.1816 & 0.96 &0.3069 & 0.0692& 0.0169& 0.0933 & 0.0968& 0.96 & 0.0732& 0.0389 \\
2 & $\beta_2$ & 0.0216 & 0.1783 & 0.1821 & 0.98 &-0.1738&0.0887& 0.0149& 0.0963 & 0.0967& 0.96 &-0.0304& 0.0461\\
\cline{2-14}
& & \multicolumn{5}{c}{$\epsilon\sim$ Normal }&&
  \multicolumn{6}{c}{ $\epsilon\sim$ Logistic} \\
& $\beta_1$ & 0.0020 & 0.1469 & 0.1443 & 0.95 &0.2015& 0.0490& 0.0155&0.2835  &0.2755 &0.98 &0.4789&0.0739 \\
& $\beta_2$ & 0.0006 & 0.1442 & 0.1431 & 0.97 &-0.1217 & 0.0600& -0.0256&0.3052  &0.2656 &0.98 &-0.2841 & 0.0993   \\
\cline{1-14}

& & \multicolumn{5}{c}{$\epsilon\sim$ Double Exponential}&&
  \multicolumn{6}{c}{$\epsilon\sim$ Extreme Value} \\
& $\beta_1$ & 0.0161 & 0.1426 & 0.1394 & 0.96 &0.1556&0.1198& 0.0087& 0.0649 & 0.0801& 0.99&-0.0620&0.2188 \\
3 & $\beta_2$ & 0.0155 & 0.1593 & 0.1384 & 0.96 &-0.0297&0.1683& 0.0032& 0.0786 & 0.0811& 0.98 &0.1327&0.3064\\
\cline{2-14}
& & \multicolumn{5}{c}{$\epsilon\sim$ Normal }&&
  \multicolumn{6}{c}{ $\epsilon\sim$ Logistic} \\
& $\beta_1$ & 0.0067 & 0.0989 & 0.0957 & 0.94 &0.0364&0.1554& 0.0354 & 0.2108 & 0.1931& 0.99 &0.1795&0.1376\\
& $\beta_2$ & 0.0017 & 0.0880 & 0.0937 & 0.99 &-0.0249&0.1484& 0.0239& 0.1986 & 0.1870&0.98 &-0.0167&0.1744 \\

\cline{1-14}

& &\multicolumn{5}{c}{$\epsilon\sim$ Double Exponential}&&
  \multicolumn{6}{c}{$\epsilon\sim$ Extreme Value} \\
& $\beta_1$ & 0.0306 &	0.0784&	0.0874 &	0.96	&0.0035&0.0746 &	0.0272&	0.0825	 &0.0858&	 0.96&-0.2608&0.0681\\

 4 & $\beta_2$ & 0.0036 &	0.0708 &	0.0887&	0.99	&-0.0072&0.0851&	0.0031	 &0.0717&	 0.0864&	0.99& 0.1630&0.0786\\
\cline{2-14}
 & &\multicolumn{5}{c}{$\epsilon\sim$ Normal }&&
\multicolumn{6}{c}{ $\epsilon\sim$ Logistic} \\
& $\beta_1$ & 0.0028&	0.0948&	0.0797	&0.96 &0.0035&0.0527&	0.0245&	0.1282&	 0.1263&	 0.98 &0.1689&0.0844\\

& $\beta_2$ & 0.0014&	0.0794&	0.0794&	0.96 &-0.0054&0.0615&	0.0105	&0.1005&	 0.1225&	 0.99&-0.0147&0.1011\\
\cline{1-14}

& & \multicolumn{5}{c}{$\epsilon\sim$ Double Exponential} &&
 \multicolumn{6}{c}{$\epsilon\sim$ Extreme Value} \\
& $\beta_1$ & 0.0070 &	0.2447&	0.2464&	0.97	&0.0995&0.0760&	0.0451 &	 0.2478&	 0.2332	 &0.97 &-0.1797&0.0712\\

 5 & $\beta_2$ & 0.0206 &	0.2971	&0.2496 &	0.96 &-0.1329&0.0843& 0.0081 &	 0.2227&	 0.2381&	 0.97&-0.0206&0.0880 \\
\cline{2-14}
& & \multicolumn{5}{c}{$\epsilon\sim$ Normal }&&
  \multicolumn{6}{c}{ $\epsilon\sim$ Logistic} \\
& $\beta_1$ & 0.0262 &	0.4074	&0.3964& 	0.96	&0.1311&0.0496&	0.0417 &	 0.2045	 &0.2083 &	0.98&-0.0349&0.0832
\\

& $\beta_2$ & 0.0402&	0.4028&	0.3907&	0.96 &-0.1645&0.0583&	0.0389 &	 0.1942&	 0.2082 &	 0.95 &0.2748&0.0952\\
\cline{1-14}
\cline{1-14}
\multicolumn{13}{l}{Note: S represents sampling scheme;  IP represents the inverse probability method. }
\end{tabular*}

%\end{center}
}
\end{table}

%\begin{center}
%\textbf{Table 2:} \textit{Simulation results with dependent $X$ and $Z$ }
%\end{center}

%\begin{table}
%\centering
%\caption{\textit{Simulation results with dependent $X$ and $Z$.}}\label{table2}
%\begin{tabular}{cccccccccc}
%\hline\hline
% & BIAS & SE & SEE & CP & &  BIAS & SE & SEE & CP \\
%\hline
%  & \multicolumn{4}{c}{$\epsilon\sim$ Double Exponential}& &
%  \multicolumn{4}{c}{$\epsilon\sim$ Extreme Value} \\
%$\beta_1$ & 0.0501 & 0.2530 & 0.2501 & 0.96 && 0.0417& 0.2045 & 0.2083& 0.98\\
%$\beta_2$ & -0.0206 & 0.2971 & 0.2496 & 0.96 && -0.0081& 0.2227 & 0.2381& 0.97\\
% \cline{1-5}\cline{7-10}
%& \multicolumn{4}{c}{$\epsilon\sim$ Normal }& &
%  \multicolumn{4}{c}{ $\epsilon\sim$ Logistic} \\
%$\beta_1$ & -0.0028 & 0.1326 & 0.1404 & 0.95 && -0.0023& 0.1915 & 0.1799& 0.92 \\
%$\beta_2$ & 0.0050 & 0.1195 & 0.1279 & 0.98 && -0.0021& 0.1945 & 0.1660& 0.95 \\
%\hline\hline
%\end{tabular}
%\end{table}

%\begin{table}
%\centering
%\caption{\textit{Results of MRC and LS with Forbes Global 2000 data.}}\label{table3}
%\begin{tabular}{c|cc|cccc}
%
%\hline\hline
%&\multicolumn{2}{c|}{MRC}&\multicolumn{4
%}{c}{Least Squares}\\
%\hline
%& $X_1$ & $X_2$ & $Intercept$ & $X_1$ & $X_2$
%& $Z$\\
%Est & 0.0209 & 0.0544 & 2.7512 & 0.0263 &0.0659 & -0.9521\\
%SE & 0.0056 & 0.1273 & 0.1366 & 0.0061 &0.1702 & 0.2768\\
%\hline\hline
%\end{tabular}
%\end{table}

\vspace{0.1in}
\end{document}